EDUCATIONAL ARTICLE  OPEN ACCESS

# Atmospheric Drag, Occultation 'N' Ionospheric Scintillation (ADONIS) mission proposal

## Alpbach Summer School 2013 Team Orange

Sebastian Hettrich[1,2], Yann Kempf[3,4,*], Nikolaos Perakis[5], Jędrzej Górski[6], Martina Edl[7], Jaroslav Urbář[8,9], Melinda Dósa[10], Francesco Gini[11], Owen W. Roberts[12], Stefan Schindler[13], Maximilian Schemmer[14], David Steenari[15], Nina Joldžić[13], Linn-Kristine Glesnes Ødegaard[16,17], David Sarria[18], Martin Volwerk[19], and Jaan Praks[20]

1. German Federal Office for Radiation Protection, Oberschleißheim, Germany
2. Meteorological Institute, Ludwig Maximilian University, Munich, Germany
3. Finnish Meteorological Institute, Helsinki, Finland
   *Corresponding author: yann.kempf@fmi.fi
4. University of Helsinki, Helsinki, Finland
5. Department of Aerospace Engineering, Technical University of Munich, Munich, Germany
6. Wroclaw University of Technology, Wroclaw, Poland
7. Karl-Franzens University, Graz, Austria
8. Faculty of Mathematics and Physics, Charles University, Prague, Czech Republic
9. Institute of Atmospheric Physics, Academy of Sciences of the Czech Republic, Prague, Czech Republic
10. Space Research Group, Eötvös Loránd University, Budapest, Hungary
11. CISAS, University of Padova, Padova, Italy
12. Department of Mathematics and Physics, Aberystwyth University, United Kingdom
13. Vienna University of Technology, Vienna, Austria
14. École Normale Supérieure de Lyon, Lyon, France
15. Division of Space Technology, Luleå University of Technology, Kiruna, Sweden
16. Birkeland Centre for Space Science, Bergen, Norway
17. Department of Physics and Technology, University of Bergen, Bergen, Norway
18. IRAP, UPS-OMP CNRS, Toulouse, France
19. Space Research Institute, Austrian Academy of Sciences, Graz, Austria
20. Aalto University, Espoo, Finland

Received 28 February 2014 / Accepted 3 November 2014

## ABSTRACT

The Atmospheric Drag, Occultation 'N' Ionospheric Scintillation mission (ADONIS) studies the dynamics of the terrestrial thermosphere and ionosphere in dependency of solar events over a full solar cycle in Low Earth Orbit (LEO). The objectives are to investigate satellite drag with in-situ measurements and the ionospheric electron density profiles with radio occultation and scintillation measurements. A constellation of two satellites provides the possibility to gain near real-time data (NRT) about ionospheric conditions over the Arctic region where current coverage is insufficient. The mission shall also provide global high-resolution data to improve assimilative ionospheric models. The low-cost constellation can be launched using a single Vega rocket and most of the instruments are already space-proven allowing for rapid development and good reliability.

From July 16 to 25, 2013, the Alpbach Summer School 2013 was organised by the Austrian Research Promotion Agency (FFG), the European Space Agency (ESA), the International Space Science Institute (ISSI) and the association of Austrian space industries Austrospace in Alpbach, Austria. During the workshop, four teams of 15 students each independently developed four different space mission proposals on the topic of "Space Weather: Science, Missions and Systems", supported by a team of tutors. The present work is based on the mission proposal that resulted from one of these teams' efforts.

## 1. Introduction

The Sun is continuously emitting a stream of charged particles and electromagnetic radiation, the so-called solar wind, into interplanetary space. The solar wind is threaded by the interplanetary magnetic field (IMF). Following an 11 year cycle, the solar activity includes typical events such as Coronal Mass Ejections (CMEs), Co-rotating Interaction Regions (CIRs) and flares, which have a significant impact on near-Earth space and the Earth's atmosphere (e.g. Gosling et al. 1990; Yermolaev et al. 2005; Alves et al. 2006, and references therin). Once these solar wind particles and electromagnetic energy reach Earth, they can affect the planet's atmosphere by thermal heating and ionisation of its upper layers (Sojka et al. 2009).

This interaction between the solar radiation and Earth's upper atmosphere forms the ionosphere, consisting of positively charged particles and free electrons. The thermosphere ranging from 80 to 350 km is formed by neutral particles, with their numbers being several orders of magnitude higher and their temperature increasing with increasing altitude (Prölss 2004). The ionosphere extends typically over an altitude from 80 to 1500 km and overlaps with the thermosphere and the exosphere beyond.

Density variations of ionised particles in the ionosphere modify electromagnetic wave propagation: waves are refracted, thus having an increased physical path length (Blaunstein & Christodoulou 2007). At times of enhanced solar activity severe changes in ionospheric and thermospheric properties have been observed, often causing perturbations of communication signals as well as of Global Navigation Satellite Systems (GNSS; e.g. Prölss 2004). The frequency-dependent refraction also causes a group delay of the refracted wave, which has to be accounted for in case of GNSS signals to avoid errors in the global positioning and to ensure the usability of High-Frequency (HF) radio signals during strong solar events (e.g. Hunsucker & Hargreaves 2003). Especially energetic events can cause the deviation of navigational systems to exceed the acceptable limits set by the United States Federal Aviation Authority (FAA; Hapgood & Thomson 2010). A famous example is the space weather event which occurred in October 2003, called the Halloween storms. Among other effects, more than 30 satellite operational anomalies were reported (Weaver et al. 2004).

Many projects to study the properties of Earth's upper atmosphere and ionosphere have been performed over decades in the Arctic region using ground-based measurements (ionosondes, radars, using GNSS signals) as well as sounding rockets (e.g. Reinisch 1986). The effects on the thermosphere and ionosphere due to space weather are still not well understood and modelling at a global scale is difficult due to a number of reasons (e.g. low coverage of model input data over the oceans, impulsive changes). Effects on the polar regions have been studied intensively and the characteristics of space weather events such as substorms (Akasofu 1964) are well known. However, an accurate forecast for the arrival time of a CME at Earth is difficult and the errors can be of several hours (e.g. Gopalswamy et al. 2001). Predicting solar flares is even more problematic, and the associated energetic X-rays travel to Earth at the speed of light, leading to ionospheric perturbations that can occur without warning. Additionally, ground-based measurements are very limited from a global perspective, in general the coverage is sporadic and inhospitable areas such as oceans and deserts are barely covered at all (see Fig. 1 by Reinisch & Galkin 2011). To improve coverage, one solution would be to deploy GNSS instruments on ocean buoys, ships and aircraft. While this would improve global coverage significantly it does not provide uniform coverage. A possible solution to both issues of global and uniform coverage of the ionosphere is to observe from space. Radio occultation measurements can provide the data required on a global scale (Schreiner et al. 1999).

Radio occultation has been used previously for meteorological applications in the neutral atmosphere to obtain atmospheric pressures and temperatures (e.g. Wickert et al. 2005; Healy & Thépaut 2006), and in the ionosphere to measure the total electron content (TEC). The Constellation Observing System for Meteorology, Ionosphere, and Climate (COSMIC) and the Challenging Minisatellite Payload (CHAMP) conducted radio occultation measurements (IJssel et al. 2005), while the Space Test Program Satellite-1 (STPSat-1) microsatellite surveyed ionospheric scintillation with its Computerized Ionospheric Tomography Receiver In Space (CITRIS; Bernhardt & Siefring 2010). Radio occultation is a feasible method for monitoring the ionosphere on a global scale and at a high resolution in time. Such data would enable the validation of existing models and the improvement of assimilative models of the ionosphere (Hajj & Romans 1998; Schreiner et al. 1999; Bilitza et al. 2011).

The heating caused by solar events leads to thermal expansion of the upper atmosphere. Heating in the auroral region is also the source of phenomena such as large-scale atmospheric gravity waves (e.g. Crowley & Williams 1987, and references therein) or perturbations in the thermospheric winds (e.g. Thuillier et al. 2005, and references therein). Low orbiting spacecraft consequently experience changes in drag due to the temporary fluctuations in density and velocity, resulting in orbit changes and increased deceleration. Common examples are the re-entry of the Skylab mission (Smith 1978) and the fast decay of the International Space Station's orbit requiring frequent altitude boosts (European Space Agency 2011). In some cases, the effect can be terminal for a spacecraft or space debris and initiate re-entry into the Earth's atmosphere. In order to better understand spacecraft orbit decay and to enable more accurate prediction for re-entry of space debris and the disposal of spacecraft, it is necessary to discern the effect that space weather has on atmospheric drag on a global scale.

The drag experienced by a spacecraft is described by (Vallado & Finkleman 2014):

$$a_d = \frac{1}{2m} \rho v^2 A c_d, \quad (1)$$

$$c_d = c_d(T_0, T_S, n_p, m_p), \quad (2)$$

where $a_d$ is the acceleration of the spacecraft due to the drag, $c_d$ the drag coefficient, $m$ the mass of the spacecraft, $A$ the spacecraft cross-section, $v$ the relative velocity with respect to the atmosphere, $n_p$ the particle number density, $m_p$ the mean molecular mass, $\rho$ the atmospheric density and $T_0$, $T_S$ the temperatures of the atmosphere (neutrals) and the spacecraft surfaces, respectively. The dependence of drag on these parameters is one of the main problems in building an accurate drag model and subsequently deriving precise atmospheric models.

Basic research on drag effects has been conducted for several decades. Jacchia (1970) published several versions of the Jacchia Reference Atmosphere model during the early 1970s including atmosphere density and temperature parameters based on satellite drag data. Gaposchkin & Coster (1988) used precision tracking data on three spherical satellites and evaluated them based on several thermospheric models, including the Jacchia models. They identified numerous uncertainties in drag modelling and the inadequacy of thermospheric models. Hedin (1987) and his group introduced the new MSIS model (Mass Spectrometer, Incoherent Scatter) using the data of the named instruments and providing atmospheric composition data. Since then, atmospheric density models have been improved (MSISE-90 model by Hedin 1991; NRLMSISE-00 model by Picone et al. 2002) which resulted in a decrease of model errors (Volkov et al. 2008). Satellite aerodynamics as well as upper atmospheric density and wind profiles have been subject to recent investigations, being the major source of error in the spacecraft and space debris orbit determination and prediction (Doornbos et al. 2005; Doornbos 2011; Koppenwallner 2011). As Vallado & Finkleman (2014) point out, currently the main difficulty in drag modelling is to identify a common approach in the determination of the satellite aerodynamic parameters and of the atmospheric properties. The different assumptions upon which the atmospheric models are based (e.g. approximations in the thermospheric density, solar flux, solar wind – atmosphere interactions), lead to solutions and models that are substantially different but equally valid. For this very reason there is a strong need for the direct measurement of drag effects – preferably at different altitudes simultaneously – instead of its derivation from models. Data of this kind contributes to precise orbit determination, mass determination and investigation of geophysical phenomena (Gaposchkin & Coster 1988).

Recently the European Space Agency's (ESA) Gravity Field and Steady-State Ocean Circulation Explorer (GOCE) satellite has also performed indirect drag measurements due to its sophisticated drag-free mode control system, at an altitude of about 250 km (Canuto 2008). In the near future, the QB50 mission, which will use a constellation of 50 CubeSats, plans to investigate the drag in the lower thermosphere over a period of 3 months (Gill et al. 2013). Although QB50 will perform drag measurements, due to the short mission duration these measurements are not comprehensive nor do they monitor the variations over a full solar cycle. To entirely grasp the connection between increase in drag and solar events, an in-depth study over the duration of a full solar cycle is required to gather enough data such that a large number of space weather events and their effects can be studied over all phases of the solar cycle.

Therefore, to improve our understanding of the dynamical behaviour of the terrestrial ionosphere and thermosphere due to changes in the solar activity over the course of a solar cycle, we propose the Atmospheric Drag, Occultation 'N' Ionospheric Scintillation mission (ADONIS). Intending to monitor ionospheric changes due to space weather, the mission will take measurements related to two phenomena: first, as the upper atmosphere is heated and perturbed, the gas dynamic drag effects increase and satellite orbits are altered. ADONIS will measure the acceleration due to drag and parameters related to it. Second, the enhanced level of ionisation changes the propagation of radio waves. Signals undergo several kinds of alteration when propagating in the ionosphere, even in geomagnetically calm periods through refraction, Faraday rotation, group delay and dispersion. Concerning radio waves used in satellite communication, Faraday rotation is negligible (a few degrees) and dispersion is weak (a dozen picoseconds per one megahertz) (Blaunstein & Christodoulou 2007). Thus, ADONIS will focus on refraction and group delay with occultation measurements. In case of a disturbance, signal amplitude and phase scintillation appears, which enables measurements of plasma inhomogeneities (e.g. Mitchell et al. 2005).

The mission will provide near real-time (NRT) ionospheric monitoring products to extend state-of-the-art space weather service capabilities as the currently used products from medium Earth orbit GNSS satellites have low performance at auroral latitudes. Ionospheric tomography provides additional monitoring services, detecting irregularities ranging up to hundreds of metres which can influence radio wave propagation through the ionosphere. Active radio beacons on the ADONIS satellites in Low Earth Orbit (LEO) will provide higher resolution of ionospheric projections with ground stations compared with the signals from standard GNSS constellations at orbits with higher apogee. Benefits have already been demonstrated for example with the Finnish ionospheric tomography chain using satellite beacons (e.g. Nygrén et al. 1996).

ADONIS is unique in that it combines both the radio occultation and scintillation measurements with drag measurements in one mission. It will address the space weather-related issues critical to LEO objects (spacecraft, space debris) which mainly are the effects on communication and orbital parameters. ADONIS consists of a constellation of two satellites on different LEOs to obtain an unprecedented global coverage over a complete solar cycle. The mission also offers flexibility since satellites could be replaced at the end of their lifetime such that ADONIS becomes a continuous space weather monitoring system. Moreover, the addition of further satellites could increase the spatial resolution, due to the simultaneous measurements covering a larger part of the ionosphere. The presence of more orbital planes would also increase the frequency with which the same atmospheric regions are covered due to the precession of the trajectory. At the same time, a better time resolution in universal time of the polar regions could be achieved. Since the satellites move in nearly polar trajectories, their orbits intersect over the polar regions, thereby producing a coverage of these areas twice during the 96 min of an orbit's duration. A larger number of satellites could therefore improve the resolution of these high latitude areas proportionally.

ADONIS is a low-cost and efficient method for monitoring the ionosphere globally and providing a valuable new source of data for assimilative models. It will be able to run continuously, improving on and complementing current efforts based on ground- and space-based GNSS TEC measurements as well as ground-based ionosonde networks.

This paper is organised as follows: In Section 2 a mission overview is given, Sections 3 and 4 describe the mission and the spacecraft designs. The development and costs are presented in Section 5, and finally the conclusions are presented in Section 6.

## 2. Mission overview

### 2.1. Mission statement

The ADONIS mission is proposed to study the dynamics of the terrestrial ionosphere and thermosphere over the duration of a full solar cycle. ADONIS shall determine the key parameters in the ionosphere and the thermosphere in relation to satellite drag and radio signal propagation. The long mission lifetime shall facilitate investigation of the effects of the variability of solar conditions on the Earth's atmosphere.

### 2.2. Objectives

The mission objectives are as follows:

*Objective 1*: Study the dynamics of the thermosphere and its effects on satellite drag in-situ, in the LEO region at 300–800 km. Current drag and atmospheric models of satellites show deviations up to 20% from the actual behaviour (Doornbos 2011; Mehta et al. 2013). This leads to satellite operators overestimating the fuel required, and to less accurate orbit predictions. Current modelling for satellite re-entry and space debris orbital evolution is not optimal for precise determination of their orbit and their re-entry position, which is essential to ensure a safe and controlled de-orbiting.

*Objective 2*: Measure the ionospheric response to space weather events in order to derive electron density and TEC maps. Global coverage is currently achieved using both ground-based and space-based GNSS TEC measurements used in products such as SWACI (http://swaciweb.dlr.de), which is part of ESA's Space Situational Awareness (SSA) programme. Additionally, ionosondes measure vertical density profiles of the lower ionosphere, providing valuable constraints to assimilative models, but their coverage is less global. The ADONIS mission shall improve the global coverage and the provision of TEC NRT data for the Arctic region.

*Objective 3*: Provide measurements relevant to satellite drag and to the ionospheric response to space weather events over a full solar cycle. A long mission lifetime allows the observation of a large number of similar solar events and the ionospheric and thermospheric response to the electromagnetic flux and particle precipitation. This yields a comprehensive dataset for statistical studies and improved modelling of drag as well as the ionospheric and thermospheric response to space weather events.

The ADONIS mission will not measure parameters such as the solar wind parameters or the X-ray flux, which directly affect the ionisation in the ionosphere. The interpretation of the ADONIS measurements in terms of correlations to space weather will rely on the network of existing and planned missions. For example, the Active Composition Explorer (ACE), the SOlar and Heliospheric Observatory (SOHO) and the Wind spacecraft are currently providing solar wind parameters from orbits around the Lagrangian point L1, the Solar TErrestrial RElations Observatory (STEREO) and the Solar Dynamics Observatory (SDO) monitor solar activity while the constellation of GOES spacecraft monitors radiation and particle fluxes. The Deep Space Climate Observatory (DSCOVR) is scheduled to be launched soon and to be placed near L1. Due to its use of a solar sail, the planned Sunjammer mission is even designed to monitor from a location upstream of L1. The Chinese KL5 mission is now in preparation with ample provision of international instruments. KL5 considers space weather monitoring at both L1 and L5 Lagrangian points.

In the highly complex atmosphere–ionosphere system it will be a challenge to integrate in-situ measurements of the thermosphere and spatially non-collocated ionospheric measurements. The mission nevertheless aims at investigating the reaction of the two regions to space weather events, and to contribute to better modelling with high-resolution data in both cases. The strength of ADONIS is that measurements of thermospheric and ionospheric parameters are carried out at the same time with high cadence, so as to identify and follow changes caused by space weather.

### 2.3. Requirements

The requirements for the space mission derived from the objectives are given in this section. The ADONIS mission shall measure the acceleration on the spacecraft, the atmospheric composition and the spacecraft temperature in order to provide key parameters for a better understanding of atmospheric drag in the high atmosphere (*Obj. 1*). In order to provide the parameters affecting telecommunications and navigation, global electron density profiles as well as the plasma parameters shall be determined (*Obj. 2*). The mission shall last at least 11 years (*Obj. 3*). A more detailed overview of the requirements follows.

*Requirement 1*: The mission shall provide in-situ measurements of the plasma and neutral densities, temperature, velocity (in the ram and transverse directions) and the spacecraft external surfaces' temperature, which are relevant to model spacecraft drag at the required altitude (Pilinski et al. 2011, 2013). The velocity measurements provide information on plasma flow in the ionosphere, and they also augment ground-based radar measurements, which provide line-of-sight ion bulk velocity either away from or towards the beam source.

The required accuracies and cadences for these parameters are:

- Plasma and neutral densities: ±5% at 1 Hz;
- Plasma and neutral temperatures: ±5% at 1 Hz;
- Velocities: ±5% at 1 Hz;
- Spacecraft temperature: 1 K at 1 Hz.

*Requirement 2*: The mission shall determine the acceleration with an accuracy of $10^{-8}$ ms$^{-2}$ and a cadence of 1 Hz. Based on the NRLMSISE-00 atmospheric model the range of expected acceleration due to drag is about $10^{-6}$–$10^{-8}$ ms$^{-2}$ (Picone et al. 2002). In order to cover small changes in the drag acceleration even at high altitudes, an accuracy of at least $10^{-8}$ ms$^{-2}$ is required. ADONIS shall measure the acceleration of the spacecraft with a cadence of 1 Hz.

*Requirement 3*: The mission shall provide a global daily coverage with a longitudinal separation lower than 15°, because *Obj. 1* and *2* require a low longitudinal separation with a short repetition time. Measurements shall be made at altitudes covering a range from 300 to 800 km encompassing both the bottom-side and the top-side F-region of the ionosphere, because drag measurements require low passes whereas ionospheric measurements are more complete from higher altitudes. This allows to cover global changes affecting satellite drag and ionospheric conditions and to provide corresponding datasets.

*Requirement 4*: The mission shall provide NRT coverage of the Arctic region (above 63° N latitude). The strongest and most recurrent ionospheric space weather effects occur at high latitudes, but ground-based measurements are sparse in these regions. A uniform spatial coverage in NRT above the Arctic region is required in addition to *Req. 3* to meet *Obj. 2*.

*Requirement 5*: The mission data shall enable the derivation of electron density profiles with altitude resolution of 1 km. The mission shall use radio occultation to derive the electron density profile. Ionospheric scintillation measurements shall also be used, to gain information about transient ionospheric phenomena such as polar cap ionisation patches and how they affect signal propagation (e.g. Zhang et al. 2013).

*Requirement 6*: The mission shall determine the magnetic field in-situ with cadence higher than 1 s and resolution of 0.5 nT for a dynamic range of ±80,000 nT. This is required to get a complete picture of magnetic field changes due to solar wind influences causing ionospheric perturbations. The magnetic field in the ionosphere can vary between magnitudes of 25,000 nT at the equator and 65,000 nT at the poles. Small perturbations of the order of nT are caused by ionospheric currents and plasma waves, these perturbations have time scales of a few seconds (Alperovich & Fedorov 2007), therefore a resolution of 0.5 nT is required at sub-second cadence.

*Requirement 7*: The mission shall operate for the duration of a full solar cycle. In order to satisfy *Obj. 3* the mission has an initially planned lifetime of 11 years.

## 3. Mission design

In order to meet the requirements mapped out in Section 2, the ADONIS mission is designed as a constellation of two identical spacecraft, A-DONIS and B-DONIS, orbiting in LEO in perpendicular orbital planes, covering altitudes of 300–800 km with constantly moving apogee. The constellation can be launched with a single launch vehicle.

### 3.1. Orbit

The final satellite configuration consists of two different elliptical orbital planes, both with an inclination of 80°, the apogee at 800 km and perigee at 300 km altitude. Figure 1 gives a schematic view of the orbits. The need for a nearly polar orbit arises from the fact that a major goal of the mission is to provide measurements at high resolution over the whole globe. An increased inclination ensures the coverage of areas at high latitudes thereby fulfilling this mission objective.

The orbit was intentionally chosen not to be sun-synchronous to enable the observation of the same location at different local times. This choice is justified by taking into account that different local times correspond to different sunlight conditions and therefore different atmospheric and ionospheric parameters covered.

The apogee altitude was limited to decrease the total ionisation dose which would be accumulated over the 11-year lifetime. In general, a higher altitude not exceeding the ionospheric boundary would be of benefit due to the increase of available measurements, but at the same time it would have to be guaranteed that the drag experienced at these high altitudes could still be measured with the on-board sensors. The perigee height was chosen to be low, since the presence of higher density layers (resulting in higher accelerations due to drag) would allow the accelerometer to perform accurate measurements over a wider range of altitudes, without increasing the drag too much thereby reducing the amount of fuel required.

The formation of the final orbit configuration was chosen such that only one launch is required, and was therefore designed with the two satellites initially being positioned in the same orbital plane in a 300 × 800 km trajectory after the launcher burn out. An impulse change is imparted on A-DONIS ($\Delta V$ = 0.14 km/s) and on B-DONIS ($\Delta V$ = 0.09 km/s) in order to achieve a circular (800 km) and elliptical (300 × 500 km) orbit, respectively. The orbital drift created by the second harmonic of the Earth's gravitational field has different values for orbits with unequal eccentricities, therefore causing a relative precession rate of the Right Ascension of the Ascending Node (RAAN) for the spacecraft with a value reaching 0.255°/day. By lowering the inclination of the orbits from (the theoretically ideal for coverage) 90° down to 80°, it is possible to take advantage of this drift and control its precession rate. The difference in RAAN becomes equal to 90°, 340 days after launch. The same $\Delta V$ as for the initial orbits are applied to A-DONIS and B-DONIS, respectively (apogee kick burn), in order to achieve two identical orbits (300 × 800 km, period of

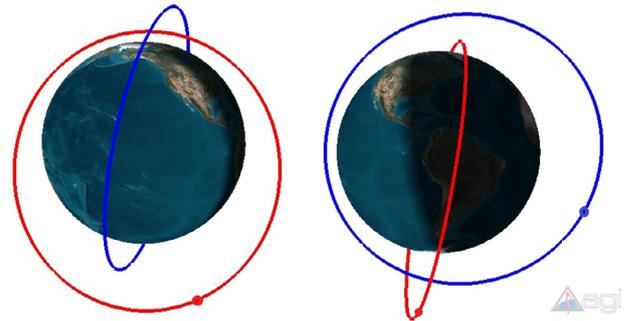

**Fig. 1.** The orbits of A-DONIS (red) and B-DONIS (blue) in their final configuration (eccentricity exaggerated).

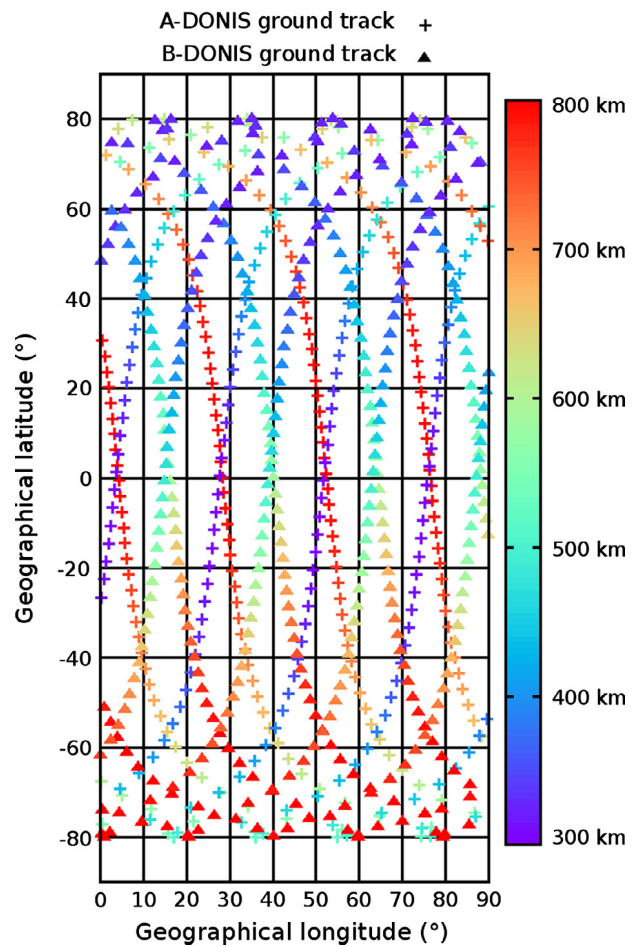

**Fig. 2.** Coverage of the ADONIS mission in one day. Longitude shown from 0 to 90°. The coverage is identical for other longitudes.

96 min), with a 90° difference in the two orbital planes (RAAN) and 90° difference in the argument of perigee (see Fig. 1). This configuration buildup allows a reduction in the amount of propellant required and the cost is lowered, since the plane change is carried out by taking advantage of the gravitational perturbations rather than an on-board propulsion system or an additional launch.

The final configuration provides the required spatial and temporal coverage, as shown in Figure 2. It is important to note that the RAAN and the argument of perigee continue to precess throughout the mission duration with the same rate for both orbits, providing measurements over a wider longitude-latitude and local time combination range. Specifically, four local times are covered in each latitude (two from each spacecraft) and the precession of the RAAN causes these local times to shift. The RAAN shift occurs with 1.3°/day, which translates to a 5.2-min shift of the local time every day. After 70 days, all local times are covered for a specific latitude and the cycle starts over, corresponding to an examination of almost 57 full coverages in local time for each latitude within the 11-year cycle. The extensive and repetitive coverage of the 24-hour cycle shall provide sufficient measurement points for a complete description of the local time effects on the spacecraft – atmosphere interaction.

Table 1. Instrument range and sensitivity corresponding to the requirements for the mission instruments. NRT: Near Real-Time. Detailed descriptions of the instruments are given in the text.

| Requirements | Range | Sensitivity, rate | Instrument name |
|---|---|---|---|
| Particle composition | 0–50 amu | 0.4 amu, 1 Hz NRT | Ion and Neutral Mass Spectrometer (INMS) |
| On-board temperature | −170 to +160 °C | 1 °C, 1 Hz | Thermistors |
| S/C acceleration | 0–5 g | $10^{-8}$ ms$^{-2}$, 1 Hz | Italian Spring Accelerometer (ISA) |
| Plasma velocity, temperature | ~0.03–1.3 eV | 16 Hz NRT | Ion Velocity Meter (IVM) |
| Plasma density | $10^9$–$10^{12}$ m$^{-3}$ | $10^9$ m$^{-3}$, 1 Hz NRT | multi-Needle Langmuir Probe (m-NLP) |
| Total electron content | 3–300 TECU | 3–5 TECU, 10 Hz | Integrated GPS Occultation Receiver (IGOR) |
| Total electron content | 3–300 TECU | 1 TECU, 10 Hz S4 | Radio tomography receiver (CITRIS) |
| Magnetic field | ±80 µT | 0.5 nT, 10 Hz NRT | Fluxgate Magnetometer (FGM) |

The mission's orbit improves on the coverage provided by the past GOCE (Drinkwater et al. 2003) and the planned QB50 (Gill et al. 2013) missions, which can also measure drag. Because of its main mission objective of measuring the Earth's gravitational potential from a very low orbit, GOCE constantly compensated the drag of the atmosphere to ensure accurate measurements of gravity. Tracking this compensation allows to study atmospheric drag, but at a lower altitude of 255 km than ADONIS. The CubeSat project QB50 is dedicated to study drag using 50 spacecraft in very low orbits, but the mission lifetime is only 3 months. There are other already launched or in preparation missions for measuring the atmospheric drag (e.g. Jasper & Kemble 2009), but they focus on lower orbits and shorter time periods. ADONIS shall operate at an altitude of 300–800 km which has not been studied yet in detail over a long period. This range of altitude surpasses the ones provided by the QB50 mission, which focusses on the lower thermosphere. With a planned duration of at least one solar cycle, the measurements of drag extend over a wide range of ionospheric excitation conditions, leading to a more complete description of the solar activity's influence on the interaction between spacecraft and atmosphere, compared to the aforementioned missions.

### 3.2. Instrumentation

Most instruments selected to fulfil the mission requirements have heritage from previous missions and are space proven. According to ESA's Strategic Readiness Levels guidelines (European Space Agency 2012), these components possess a Technical Readiness Level (TRL) between 7 and 9, which minimises the development costs. A notable exception is the Italian Spring Accelerometer (ISA), which is however already undergoing qualification testing on ground and in-flight (see below), meaning that its TRL will increase sufficiently by the time the ADONIS mission will be built. The list of instrument ranges and resolutions is given in Table 1, and further technical details are given in Table 4.

The Ion and Neutral Mass Spectrometer (INMS) contributes to *Req. 1* (see Sect. 2.3). It measures the mass spectrum of low-mass ionised and neutral species. Neutral compounds are important for drag measurement since their density dominates the ionised particle density, and the change in total composition is important for ionospheric monitoring. The instrumental design is intended to draw on the experience gained in the QB50 project (Gill et al. 2013), in which a similar instrument will be used in a large constellation of CubeSats. The range and sensitivity of the measurements are adjusted to our mission requirements.

The ion velocity meter (IVM) is chosen to meet *Req. 1* and consists of two instruments: a retarding potential analyser (RPA) which measures the energy distribution and an ion drift meter (IDM) which measures the arrival angles of the particles. Together these instruments provide the temperature and velocity of protons and other ions (Heelis & Hanson 1998). These instruments face the direction of the spacecraft velocity (the ram direction) and can resolve the full velocity vector in the ionosphere. It should be noted that this is under the assumption that the spacecraft's velocity is larger than the plasma bulk velocity, an assumption that is often justified in the ionosphere at altitudes above 250 km where ADONIS will operate (Heelis & Hanson 1998). The combined instrument has a rich heritage and has been flown in the relevant environment on several spacecraft such as the ones below. The ROCSAT-1/FORMOSAT-1 spacecraft, which was in a circular orbit at an altitude of 600 km, included such an instrument in its IPEI (Ionospheric Plasma and Electrodynamics Instrument) instrument (Yeh et al. 1999). The CINDI (Coupled Ion-Neutral Dynamics Investigation) mission also included such an instrument which operated at altitudes of 400–860 km. The instrument is also included on the planned ICON (Ionospheric Connection Explorer) mission. Since the instrument is proven at all altitudes where ADONIS will operate, it has a high TRL of 8–9.

The multi-needle Langmuir probes (m-NLP) are used to measure the electron density and contribute to *Req. 5*. They also provide in-situ data complementary to the occultation and scintillation measurements. This instrument has been developed and tested on a sounding rocket as reported by Bekkeng et al. (2010), thus it has a TRL of 8–9.

Radio occultation measurements shall satisfy *Req. 4* and *Req. 5*. With the Integrated GPS Occultation Receiver (IGOR) instrument, A-DONIS and B-DONIS receive GNSS signals which get Doppler-shifted while gaining frequency-dependent (dispersive) ionospheric delay because of refraction. By measuring this frequency shift – using both L1 and L2 signals – and comparing it with the non-occulted theoretical Doppler-shift, the angle of refraction and the refractive index can be derived for one assumed tangential point. From the refractive index slant TEC values are calculated for a given height. As the occultation is rising (or setting) the IGOR instrument makes 10 measurements per second, so that a vertical electron density profile for a given latitude and longitude can be derived. IGOR has an accuracy of 3–5 TEC Units (TECU). IGOR is operational on the COSMIC/Formosat-3 constellation, thus it has a TRL of 9. Its 0.5 m three-dimensional root-mean-square (rms) position accuracy and 0.2 mm/s orbital velocity accuracy are provided by the instrument itself (Rocken et al. 2000). Electron density values derived from the measurements on the required orbit extend the high-resolution ionospheric sampling of regions with sparse coverage (Arctic regions, oceans and southern hemisphere). The development of the engineering model of the Tri-GNSS (TriG) receiver (GALILEO, GPS and GLONASS) is finished, but its 50 W power requirement is too high for a trade-off providing 500 additional radio occultations daily (Broadreach Engineering 2014), hence the current instrument choice.

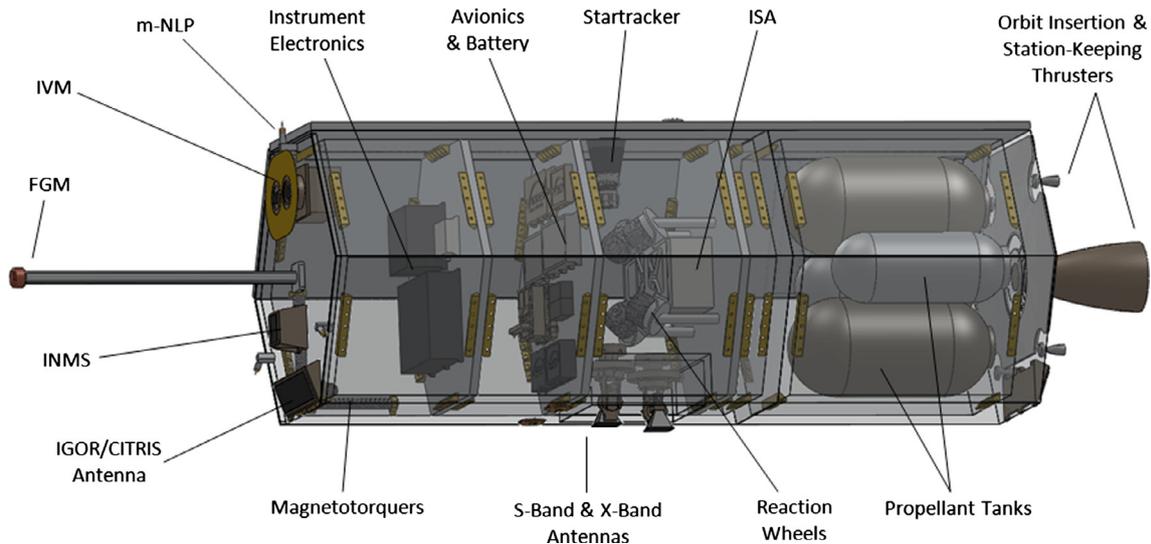

**Fig. 3.** ADONIS satellite preliminary layout.

The Computerised Ionospheric Tomography Receiver in Space (CITRIS) shall also satisfy *Req. 4* and *Req. 5*. It is designed to map local inhomogeneities in the ionosphere causing scintillations in amplitude and phase of the signals. The satellites will be using the CITRIS receiver for short-range inter-satellite scintillation measurements from the CERTO TBB radio beacons on the COSMIC satellites, as well as measurements from the over 50 active DORIS ground stations. Simultaneous integral measurements of Doppler and phase shifts are carried out, preferably along many ray paths and many different angles. These measurements are inverted and estimates for fluctuations of TEC are carried out (Howe et al. 1998). The measurement principle is similar to that of GNSS, but the ratio of the frequencies used is higher (in the case of the ground stations 5.1, other LEO satellites 2.6, while for GPS only 1.3, Nesterov & Kunitsyn 2011), and the ray paths are shorter. This tomographic method enables the derivation of TEC and plasma inhomogeneities with an accuracy of 1 TECU and 10 Hz sampling to determine the S4 scintillation index. This index quantifies the amplitude variance of the signal and is defined as the standard deviation of the normalised signal intensity over a given time. Thus, ionospheric tomography by LEO spacecraft is able to provide a higher resolution for mapping scintillation (a few up to hundreds of metres).

Scintillation is not monitored by any operational spacecraft at the moment. At scales of metres to tens and hundreds of metres, the ionosphere changes in minutes, thus rapid sampling is important. TEC data in the Arctic region shall be directly downlinked to the Svalbard ground station, providing NRT ionospheric monitoring products valuable for assimilative modelling.

Thermistors measure the surface temperature of the spacecraft, which is one of several parameters influencing the drag coefficient. Thus this measurement is needed to fulfil *Req. 1*. The solar panels are already equipped with thermistors. Additionally, there are thermistors on the front and back surfaces. Thermistors have a TRL of 9.

The Fluxgate Magnetometer's (FGM) heritage from the Cluster and THEMIS missions (Balogh et al. 1993; Auster et al. 2008) ensures a TRL of 9. It measures the changes of the magnetic field components with the range and resolution specified in Table 1 to comply with *Req. 7*. The sensor is positioned at the tip of a short 1.0 m boom to minimise its influence on the spacecraft drag while yielding reasonable magnetic field measurements. Contamination mitigation techniques developed for non-electromagnetically clean missions carrying a magnetometer will be used and can be accommodated on the spacecraft platform.

The Italian Spring Accelerometer (ISA) is scheduled to fly on BepiColombo in 2016, thus its current TRL is 4–6 but will increase by the time ADONIS is built. In order to achieve the desired accuracy of $10^{-8}$ ms$^{-2}$ with a sampling rate of 1 Hz, customisation of the ISA is necessary to satisfy *Req. 2* (Iafolla et al. 2011). The ISA is very sensitive to temperature changes and so it is covered by a thermal system to keep the temperature stable (Iafolla & Nozzoli 2001).

The preliminary layout of the spacecraft including the instruments is presented in Figure 3.

### 3.3. Launcher

The satellites are launched with an Arianespace Vega launcher. Vega has a lift-off mass of 137 tonnes and is able to carry up to 1.5 tonnes of payload to an 800 km circular orbit (Arianespace 2006). The ADONIS mission has a total payload mass of 1.1 tonnes. This leaves a margin of 400 kg for the dual launch adapter, saving costs by using the most cost-efficient rocket available to put the satellites into the required orbit. The launch site will be at the Guiana Space Centre, Kourou, French Guiana.

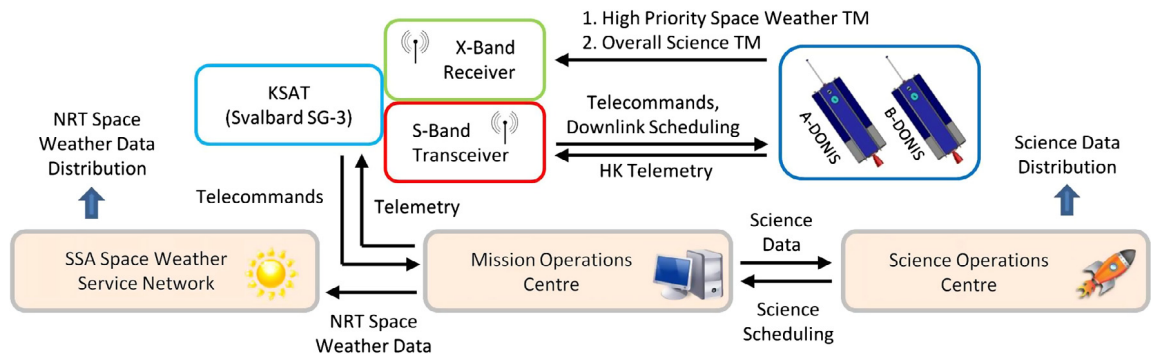

**Fig. 4.** Ground segment overview. TM: telemetry, HK: housekeeping.

### 3.4. Ground segment

The ground segment of ADONIS consists of a ground station (GS), Mission Operations Centre (MOC), Science Operations Centre (SOC) and space weather services. An overview of the mission ground segment is shown in Figure 4.

The ground station selected for the mission is the Svalbard SG-3 ground station (latitude 78° N), owned and operated by the Norwegian company Kongsberg Satellite Services AS (KSAT). It uses S-Band for downlinking housekeeping telemetry and uploading telecommands to the space segment. Science and operational telemetry data is received using X-band.

The ground station location was picked due to its position close to the North Pole, mainly for good ground coverage of satellite passes and secondly to have short time delay for downlinking the NRT space weather data in the Arctic region.

The number of ground passes changes due to the precession of the satellite orbital planes around the polar axis, which has a period of about two-thirds of a year. In the best-case scenario (when the intersection of the orbital planes is closest to the ground station) the ground station has a coverage of >95% of orbits of both satellites.

In the worst-case scenario (intersection furthest away) there is a maximum of four consecutive uncovered orbits for one of the satellites and an average of 13.5 out of 15.15 orbits per day (85%). The other satellite still remains covered at the maximum >95%. This relationship alternates between the two satellites, depending on the phase of the precession.

The average ground pass time is 10.5 min per orbit and the worst-case (usable) ground pass is 4 min per orbit, resulting in a tracking requirement of about 4 h/day for both satellites together.

The order of downlinking operational and science telemetry data is based on a combination of priority schemes and schedules. NRT data always has the highest priority (unless explicitly decided otherwise by the operations team).

The Mission Operations Centre (MOC) is responsible for monitoring and maintaining the flight critical systems of the space segment; performing orbital maintenance manoeuvres, providing interfaces to and from the Science Operations Centre (SOC) for science data, scheduling and providing data to the space weather services. Due to the use of a commercial SSTL satellite platform (see Sect. 4), standard SSTL ground station systems for ground control software and hardware are used.

The Science Operations Centre (SOC) is responsible for scheduling the science measurements, downlink schedules and priorities, as well as providing support to the MOC for instrument calibrations and maintenance. The on-board scientific instruments generate up to 350 Mbit of data per orbit during nominal operations. In the worst-case scenario (four consecutive passes without coverage), 1750 Mbit of data is produced. The worst-case downlink capacity for a pass with ground station coverage is 4500 Mbit, giving a worst-case science data downlink margin of 2750 Mbit. This also means that in case bad ionospheric conditions prevent nominal telemetry, an additional seven orbits' worth of data can be stored on board and still downlinked in one pass afterwards. The total generated scientific data during the mission is roughly 20 Tbit.

### 3.5. Disposal

ESA requires that satellites occupying LEO regions are removed and disposed of no later than 25 years after the end of the mission (European Space Agency 2008). The mission is structured so that the orbital decay is part of the scientific phase, and allows to investigate the drag at about 300 km with different perigee velocities (circularisation phase) as well as to study the drag below 300 km until the re-entry of the satellites (spiralisation phase).

To avoid the risks of an uncontrolled re-entry and because, as with the GOCE mission, the sturdy framework of the accelerometer is likely to survive re-entry in sizeable pieces (European Space Agency 2013), the mission lifetime of ADONIS shall end with a controlled re-entry using thrusters over unpopulated areas.

## 4. Spacecraft design

The ADONIS mission uses two identical spacecraft based on the commercial SSTL-300 platform (list price 23 M€), with a customised structure to meet the scientific requirements for both the drag (*Req. 1–Req. 3*) and the ionospheric measurements (*Req. 1*, *Req. 3–Req. 6*). Customisation of the satellite platform also increases the lifetime to cover a full solar cycle (*Req. 7*), ensuring the spacecraft remain operational in their environment also during space weather events.

To satisfy *Req. 1*, each spacecraft has a minimised frontal area (0.8 m$^2$) and a simple hexagonal shape with one 1.0 m boom deployed in the ram direction. The main structure is a lightweight aluminium skin frame with aluminium skinned honeycomb panels. The mass for each spacecraft is 560 kg including payload and fuel for orbit insertion, station keeping and manoeuvres satisfying *Req. 7*. Figure 3 shows a preliminary design of the spacecraft and the position of the instruments and substructures. The instruments are arranged in such a way that they do not interfere with each other.

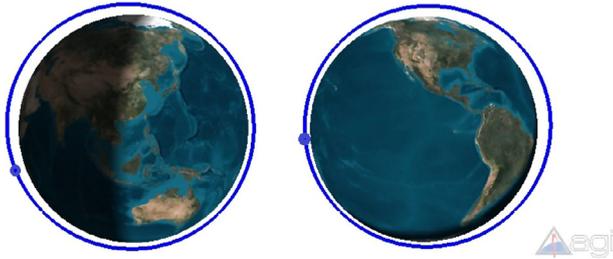

**Fig. 5.** Cold-case orbit scenario with the Sun from the right (left panel) and hot case orbit scenario with the Sun from the front (right panel).

### 4.1. Power subsystem

The power subsystem design depends on the chosen orbit, the nadir-pointing attitude and the drag measurements which are carried out. The orbital plane of the satellite slowly precesses with respect to the incoming solar radiation, which means that during its entire lifetime, each spacecraft has the incoming solar radiation on all its surfaces except the bottom. The spacecraft experience significant drag in orbit, for these reasons only body-mounted solar cells on the top and lateral panels are selected instead of deployable ones. The solar panel area is designed for the worst- and cold-case scenario (see Fig. 5 left), where the solar panels are in eclipse for half the orbit ($T$ = 48 min).

Spectrolab 29.5% NeXt Triple Junction (XTJ) GaInP2/GaAs/Ge solar cells are selected, providing a high specific power $P_0$ = 398 W/m$^2$ with a yearly degradation of 2.75%. Considering the losses in the efficiencies (assembly 5%, shadowing 5%, temperature 15%) and the total life degradation (30%) the total designed exposed area is $A_{SP}$ = 2.5 m$^2$. One solar panel on the top (0.56 × 2.6 m$^2$) and two on the top lateral panels (0.56 × 1.8 m$^2$) provide the necessary power, in the worst-case scenario, at the end of life.

Lithium ion batteries are chosen to cover the power requirements during the eclipse periods, with a capacity of 38 Ah, maximum current of 38 A and nominal voltage of 3.6 V. One single battery provides 90 W with a mass below 1 kg (including a margin of 20%). The depth of discharge is about 75%.

The power budget is included in Table 4.

### 4.2. Thermal design

The thermal design of the satellites is conceived to cope with the hot and cold orbit scenarios, which are presented in Figure 5. Thermal insulating material (MultiLayer Insulation – MLI) is designed to cover all the exposed areas of the lateral panels of the spacecraft, beneath the solar panels, in order to thermally decouple them from the satellite body.

The radiators are designed based on the radiative thermal exchange between the spacecraft, the incoming solar flux and the Earth's radiation flux along the orbit. The following optical properties are chosen for the radiators: emissivity $\varepsilon$ = 0.8 and absorption coefficient $\alpha$ = 0.1.

Due to the varying orientation of the spacecraft with respect to the solar radiation expected during the satellite's lifetime, multiple radiators covered with louvres are used. The design of the radiators is done for the hot-case scenario (see Fig. 5 right) during which the satellite is always exposed to the solar radiation (quasi dawn-dusk orbit). The radiators' dimensions are determined based on a trade-off between the excess of power produced by the lateral solar panels and the satellite total length (length of the lateral solar panels and lateral radiators). The area of each radiator is 0.4 m$^2$ including a 20% margin. The solar panels, since decoupled from the spacecraft's body, reach an equilibrium temperature during the orbit. The loss of efficiency due to off-nominal operative temperature of the solar panels is taken into account in the solar panel design.

In addition to the passive one, an active thermal control is installed in order to transfer the heat from the internal components and instrumentation towards the radiators. The active thermal control is mainly necessary during the hot case (quasi dawn-dusk orbit) during which extra power is produced.

During the cold-case scenario (in a quasi noon-midnight orbit with the longest period of eclipse, see Fig. 5 left), it is possible to reduce the power dissipation from the radiators by reducing or turning off the heat that reaches the radiators (mainly due to electrical power) or even reducing the total exposed area of the radiators with the louvres. During the worst conditions of the cold-case scenario (maximum eclipse of 34 min) heaters will provide the necessary heat to the sensitive on-board instrumentation using the batteries (20 W at worst).

### 4.3. Attitude and Orbital Control System (AOCS)

The AOCS of the spacecraft is constrained by the accuracy requirements of the payload instruments. From the AOCS point of view *Req. 1* is best met by providing constant cross-sectional area with respect to the flight vector. This means that the spacecraft is 3-axis stabilised in order to ensure identical aerodynamic conditions during the drag measurements.

In order to stabilise all three rotational axes, four reaction wheels in a pyramidal configuration will be used. The desaturation of the wheels takes place by the usage of three magnetorquers and has to be performed every 36 h, in the worst-case simulated, which includes a constant external torque around the pitch angle.

For the attitude determination, the system is equipped with 2-axis Sun sensors, one star tracker and a 3-axis magnetometer, which provides the data for the magnetic coils. The measurement of the angular velocity is carried out by a laser gyroscope. In order to make the communication with the ground station feasible, the antennas will be nadir-pointing during the passages over the ground station. For these reasons, the resulting attitude shall always be achieved by yaw steering nadir pointing mode, in order to avoid drag on the side areas of the spacecraft. The centre of mass has been chosen to be behind the centre of volume, to provide aerodynamic stability due to the restoring moment induced whenever a misalignment between the velocity vector and the vector normal to the spacecraft front face occurs.

### 4.4. Propulsion system

Both satellites have a propulsion system which is used to perform the orbital manoeuvres and corrections during the mission's lifetime. Each spacecraft needs to carry out an impulsive manoeuvre in order to change its orbital elements during the constellation build up procedure. For this reason a 100 N bi-propellant engine using monomethylhydrazine (MMH) and nitrogen tetroxide (NTO) is installed on board (specific impulse Isp = 300 s). Apart from the orbital injection, it is very important to ensure that the propulsion system is fit to compensate for the drag deceleration which the spacecraft experiences while flying in LEO. Therefore, four additional propulsion engines have been added to each satellite (10 N, Isp = 300 s) using the same propellant as the larger engine.

Table 2. $\Delta V$ and propellant budget for the ADONIS mission.

| Spacecraft | $\Delta V$ (m/s) | | Propellant mass (kg) | | Margin (%) |
|---|---|---|---|---|---|
| | A | B | A | B | |
| Injection | 280 | 185 | 58 | 32 | 25 |
| Transit corrections | 5 | 100 | 2 | 28 | 35 |
| Orbit corrections | 500 | 500 | 75 | 75 | 40 |
| Avoidances | 100 | 100 | 15 | 15 | 10 |
| Total | 885 | 885 | 150 | 150 | 10 |

Table 3. Telecommunication system summary.

| Band | X | S (down) | S (up) |
|---|---|---|---|
| Data rate | 105 Mbit/s | 38.4 kbit/s | 19.2 kbit/s |
| Frequency | 8.5 GHz | 2.2 GHz | 2.1 GHz |
| Transmission power | 5.0 W | 0.5 W | ~15 W |
| Transmission antenna | 10 cm horn | 8 cm patch | 13 m dish |
| Reception antenna | 13 m dish | 13 m dish | 8 cm dish |
| Margin | 6.0 dB | 15.4 dB | 40.6 dB |

The total $\Delta V$ change due to drag over the period of 11 years has been simulated (with overestimated solar flux and geomagnetic activity) and was used to derive the needed propellant mass for the orbital correction. The total $\Delta V$ budget and propellant needs are shown in Table 2. The propellant needed for counteracting the accumulated drag deceleration played an important role in the decision of the perigee altitude, due to the extreme increase in needed propellant mass for lower altitudes.

### 4.5. On-Board Computer (OBC) and On-Board Data Handling (OBDH)

On-board data handling and monitoring functions are provided by two redundant SSTL OBC750 on-board computers (OBC). A real-time operating system is used to support the SSTL standard spacecraft on-board software which controls and monitors the on-board systems.

A dual-redundant controller area network (CAN) bus provides communication between the subsystems and the OBC.

The control algorithm, data gathering of analogue sensors and control of actuators are provided by the ADCS (Attitude Determination and Control Subsystem) which runs on the OBC. The ADCS modules also include the interfaces for the CAN bus and analogue sensors and actuators. On-board time is provided by the GPS receivers.

The on-board science data storage is provided by a SSTL High Speed Data Recorder (HSDR), with 128 Gbit storage capacity. The HSDR also provides the interfaces between the platform and the scientific payloads, utilising the internal Low-Voltage Differential Signalling (LVDS) drivers in the HSDR for redundant 10 Mbit/s SpaceWire links to each instrument (European Cooperation for Space Standardization 2008).

### 4.6. Telecommunications

The telecommunication system uses a combination of S-band and X-band links. S-band will be used for receiving telecommands and relaying housekeeping and control telemetry to the ground. X-band will be used for downlinking the science telemetry. Table 3 gives a summary of the typical telecommunication system of the SSTL-300 commercial platform. The data rate budget is given in Table 4.

The on-board telecommunication system of the ADONIS spacecraft includes the following SSTL products: XTx400 X-Band Transmitter, S-Band Uplink Receiver and S-Band Downlink Transmitter. For the S-band systems, two opposite-facing SSTL S-Band Patch Antennas will be used (for a near spherical gain pattern), while for the X-band transmitter a SSTL Antenna-Pointing-Mechanism will be used. All antennas are redundant. The selection of the telecommunication hardware was driven by the compatibility with the SSTL platform and its heritage (Brenchley et al. 2012).

The link margins were calculated using a slant range of 2200 km from a tracking ground station 13 m dish antenna, located close to Longyearbyen on Svalbard with S-Band equivalent isotropically radiated power (EIRP) of 98 dBm, S-Band G/T of 23 dB/K and X-Band G/T of 32 dB/K (European Space Operations Centre 2008).

## 5. Development and cost

### 5.1. Total cost

ADONIS is designed to be a two-spacecraft mission as a result of the trade-off between global coverage and cost efficiency. Due to the identical design of A-DONIS and B-DONIS, the development cost due to the customisation of the satellite bus and its components turns out with estimated 60 M€ at worst-case to be relatively small and has to be spent only once.

**Table 4.** Size, power, data rate and mass of subsystems. Values given with 20% margin (10% for fuel and avionics).

|           |                | Size (cm)                   | Power (W)     | Data (bps) | Mass (kg) |
|-----------|----------------|-----------------------------|---------------|------------|-----------|
| Bus       | Structure      | 110 × 110 × 100             | N/A           | N/A        | 325       |
|           | Avionics       | 35 × 25 × 50                | 40 (61 peak)  | N/A        | 12        |
|           | Communication  | 35 × 25 × 50                | 15 (50 peak)  | N/A        | 10        |
|           | Bus Total      | N/A                         | 55 (111 peak) | N/A        | 347       |
| Payload   | INMS           | 10 × 10 × 10                | 3             | 2048       | 3.6       |
|           | IVM            | 25 × 12 × 9                 | 3             | 2000       | 2.6       |
|           | m-NLP          | 10 × 7.5 × 5                | 3.5           | 1900       | 0.3       |
|           | Thermistors    | 3.3 × 0.066 × 0.066         | 0.01          | 96         | 0.036     |
|           | CITRIS         | 40 × 31 × 12                | 12.3          | 15,000     | 5.4       |
|           | IGOR           | 21.8 × 24 × 14.4            | 22            | 20,000     | 6.96      |
|           | ISA            | 3.1 × 1.7 × 1.3             | 12.1          | 9600       | 9.78      |
|           | FGM            | 10 × 10 × 10                | 0.8           | 400        | 1.8       |
|           | Boom           | 100                         | N/A           | N/A        | 3         |
|           | Payload total  | N/A                         | 57            | 50 k       | 35        |
|           | Total dry mass |                             |               |            | 382       |
|           | Fuel           |                             |               |            | 165       |
|           | Total (wet mass)| N/A                        | 112 (168 peak)| 50 k       | 547       |

**Table 5.** ADONIS mission cost summary (in M€).

| Item             | Cost  | Amount | Total cost |
|------------------|-------|--------|------------|
| Vega launcher    | 35    | 1      | 35         |
| SSTL-300 bus     | 25    | 2      | 50         |
| Customisation    | 60    | 1      | 60         |
| Propulsion       | 17.5  | 2      | 35         |
| Full payload     | 25    | 2      | 50         |
| Ground operations| 45    | 1      | 45         |
| Mission cost     |       |        | 275        |

Due to the replicability of the spacecraft any follow-up spacecraft can be built more time- and cost-efficiently after the development of the initial prototype. The mission concept is aiming for maximal performance at minimal cost to meet all its objectives. Therefore the mission can also be easily expanded with further spacecraft in the constellation or extended in time by launching replacements.

The total cost for payload, launcher, ground operations and additional infrastructure is shown in Table 5. The expected ground operation cost is 4.05 M€/year. The satellite tracking cost was calculated with 164€/h of tracking and an additional 59€ per satellite pass with 2.2 h/day and 4200 passes/year results in 750 k€/year. The Mission Control cost is estimated by allocating 600 k€ per 24/7 operation position. With 2–3 operators and additional cost, an overall Mission Control cost of 2 M€ is expected. Science operations fall with 1 M€ for 3–4 employees into the budget, while NRT operations require one position and hardware cost of 300 k€/yr.

### 5.2. Descoping options

During the cost estimation process, the following two descoping options were identified, such that most of the mission objectives are still met.

– Use of a single spacecraft: By using one single spacecraft, the overall mission cost would decrease to 165 M€. This decreases the resolution and coverage area by half, and increases the time it takes to produce global maps, jeopardising *Req. 3* and *Req. 4*.

– Decrease of the mission duration: Decreasing the mission duration to 5 years would result in lower operational costs, therefore summing up the total costs to 251 M€ but preventing completely the fulfilment of *Obj. 3*. In case this option is applied, specific launch windows are required to launch the mission in the rising phase of the solar cycle. Since the cost savings of 24 M€ are small compared to the full mission budget, this option appears less cost efficient.

### 5.3. Mission timeline

For operational and scientific reasons, the optimal launch window is at the beginning of a solar cycle. This ensures the low solar activity at the beginning of the mission, which is optimal for instrument calibration. The drag at the solar minimum is expected to be the lowest leading to the opportunity to observe the increase with solar activity.

The next solar cycle is expected to start around 2019, thus the ideal launch period would be at that time. However, due to the prospective mission operation phase of a full solar cycle, it is not a strong requirement to launch the mission at the start of a solar cycle in order to measure drag and radio occultation in dependency of the solar maximum.

After the launch a system check and instrument calibration phase of almost one year (340 days) will commence, before the operational phase starts once the orbital planes of the two spacecraft reach the required 90° angle. At the end of the spacecraft operational lifetime, the remaining fuel will be used to increase the eccentricity of the orbits, ensuring a controlled downward spiralling of the ADONIS spacecraft, concluding with a controlled re-entry above uninhabited regions.

### 5.4. Risks

The ADONIS mission does not evidence higher risk than an average LEO mission. Of particular interest are the satellite bus customisation, the ISA instrument which has not yet been space-proven and space weather exposure.

For the bus customisation, the main risks appear in the interaction with the space environment and in combination with other subsystems. An exchange of the proposed bus system to another would increase the overall costs without significant effects in risk prevention.

For the ISA the use of different standard instruments can be considered to mitigate risks but as it is planned to fly with the BepiColombo mission its TRL will have increased sufficiently by the time ADONIS is built.

The planned bus customisation includes sufficient shielding and hardening of all components to make the spacecraft resilient to the space weather events they will be exposed to during the full mission time. Increased drag effects have been included in the propellant budget with a sufficient margin to ensure the orbit is kept for the mission duration. In case of strong perturbations of the telemetry link the spacecraft can miss 11 passes (about 16 h) as explained in Section 3. If perturbations last longer, alternate ground stations can be considered at lower latitudes. The KSAT ground station in Tromsø, Norway (latitude 69° N) is interoperable with SSTL systems but still in the auroral region and hence subject to strong ionospheric perturbations. SSTL operates a ground station in Guildford, UK (latitude 51° N) but due to its much lower latitude the coverage would be low. These options would be considered only when the margin of 11 missed passes is not enough, as using supplementary ground stations causes extra costs.

Additionally to these particular risks, launch failures were identified as risky. However since this risk is part of every space mission and due to the use of the standard launcher Vega, this is not in the range of risk mitigation in this work.

## 6. Conclusions

The ADONIS mission proposed in this paper is a constellation of two identical satellites (A-DONIS, B-DONIS) designed to study the drag on satellites and to provide a global monitoring of the ionosphere through radio occultation and scintillation measurements over the next solar cycle. The long mission timeline ensures an improvement in our understanding of how the drag on satellites and ionospheric properties are changing at times of enhanced solar activity. The uniqueness of the proposal lies in the combination of the drag and radio wave propagation studies in a single low-cost mission despite its total duration of a full solar cycle.

Better knowledge of the drag behaviour during times of enhanced solar activity will allow a more accurate estimation of the fuel needed for satellites, thus lowering the cost of future missions. Applications also encompass the forecast of the re-entry of space debris or spacecraft to be disposed of.

The interaction between the solar wind and radiation with the Earth is causing a variety of disturbances in the ionosphere influencing satellite radio signals at LEO as well as GNSS signals. Global monitoring of the ionosphere using radio occultation and scintillation measurements over a full solar cycle will lead to a more accurate determination of the signal refraction in real-time, to the valuable provision of a new source of data to assimilative models and will also contribute to the development of predictive models.

ADONIS is a low-cost mission using a single Vega rocket as launch vehicle. The satellites carry an identical set of eight instruments in order to measure the drag and ionospheric properties. The orbital configuration optimises the coverage in latitude, longitude and altitude while enabling the provision of near real-time data through the use of an Arctic ground station. The flexibility of the concept allows easy extensions of the mission, either by adding further pairs of satellites to improve the cadence of the coverage of the full globe or by replacing ageing satellites to ensure longer-term coverage than the initial mission.

*Acknowledgements.* Team Orange wants to thank our tutors Jaan Praks and Martin Volwerk for their support, dedication and contributions to our mission development. We also acknowledge in-depth discussions with and reviews from A. Balogh, V. Bothmer, Caspar, J. Eastwood, C. Erd, P. Falkner, A. de Groof, M. Hallmann, B. Lavraud, J-P. Luntama, D. Moura, M. Palmroth, A. Valavanoglou and A. Veronig throughout the development process during and after the summer school. We are grateful to both Referees for their detailed reviews and suggestions which helped greatly improve this work.

The editor thanks Mike Hapgood and Rajagopal Sridharan for their assistance in evaluating this paper.